\documentstyle[epsfig]{aipproc}

\newcommand\jhep[3]{{\it J. High Energy Phys. }{\bf #1} (#2) #3}
 
\newcommand\plb[3]{{\it Phys. Lett. }{\bf B #1} (#2) #3} 
\newcommand\prl[3]{{\it Phys. Rev. Lett. }{\bf #1} (#2) #3} 
\newcommand\prd[3]{{\it Phys. Rev. }{\bf D #1} (#2) #3}
\newcommand\prep[3]{{\it Phys. Rept. }{\bf #1} (#2) #3}

\newcommand\hepph[1]{{\tt hep-ph/ #1}}

\long\def\comment#1{}
\def\VEV#1{{\left\langle #1 \right\rangle}}

\begin{document}
\title{Indirect Detection of Neutralino Annihilation
from Three-body Channels}
\author{Xuelei Chen and Marc Kamionkowski} 
\address{Department of Physics, Columbia University,\\
538 West 120th Street New York, NY~~10027, USA\\ 
\vspace{0.3cm}
{\tt CU-TP-933, CAL-678}}
\maketitle

\section{Introduction}

The neutralino is one of the most promising dark 
matter candidates\cite{report}.
It could be detected indirectly by observation of energetic neutrinos 
from neutralino annihilation in the Sun and/or Earth.
Energetic neutrinos are produced by decays of neutralino 
annihilation products.
These neutrinos can be detected by neutrino detectors such as AMANDA and 
super-Kamiokande \cite{energeticneutrinos}, which observe the  upward muons
produced by the charged-current interactions in the rock below the detector.
For the models which can be tested by the current or next generation
of detectors, an equilibrium of accumulation and
annihilation is reached, so the annihilation rate is determined by the
capture rate in the Sun or Earth. The event rate 
is proportional to the second moment of the 
neutrino energy spectrum, so it is this neutrino energy
moment weighted by the corresponding branching ratio that determines the
detection rate.

By now, the cross sections for annihilation have been calculated
for all two-body final states that arise at tree level.
Roughly speaking, among the two-body channels the $b\bar{b}$ and
$\tau^{+}\tau^{-}$ 
final states usually dominate for $m_\chi < m_W$.  
Neutralinos that are mostly higgsino annihilate
primarily to gauge bosons if $m_\chi>m_W$,  because there is no
$s$-wave suppression mechanism for this channel.
Neutralinos that are mostly gaugino continue to annihilate primarily
to  $b\bar{b}$ pairs until the neutralino mass exceeds the top-quark mass,
after which the $t\bar t$ final state dominates, as the cross section for
annihilation to fermions is proportional to the square of the fermion mass.  

Three-body final states arise only at higher order in
perturbation theory and are therefore usually negligible.
However, some two-body channels 
easily dominate the cross section when they are open because of 
their large couplings; for example the $W^{+} W^{-}$ for the higgsinos and
$t \bar{t}$ for gauginos.
This suggests that their corresponding three-body 
final states can be important just below these thresholds. 
Moreover, the neutrinos produced in these three-body final states are 
generally much more energetic than those produced in $b$ and $\tau$ decays, 
Recently, we calculated the $s$-wave cross section for the processes
$\chi \chi \to W^{+} W^{-*} \to W f \bar{f'}$ and 
$\chi \chi \rightarrow t\bar t^* \rightarrow t W^{-} \bar{b}$, 
and their charge conjugates 
in the $v_{\rm rel} \to 0$ limit \cite{jheppaper}. 

\section{Calculation}

\begin{figure}[t]
\begin{minipage}[b]{0.45\linewidth}
\centering\epsfig{file=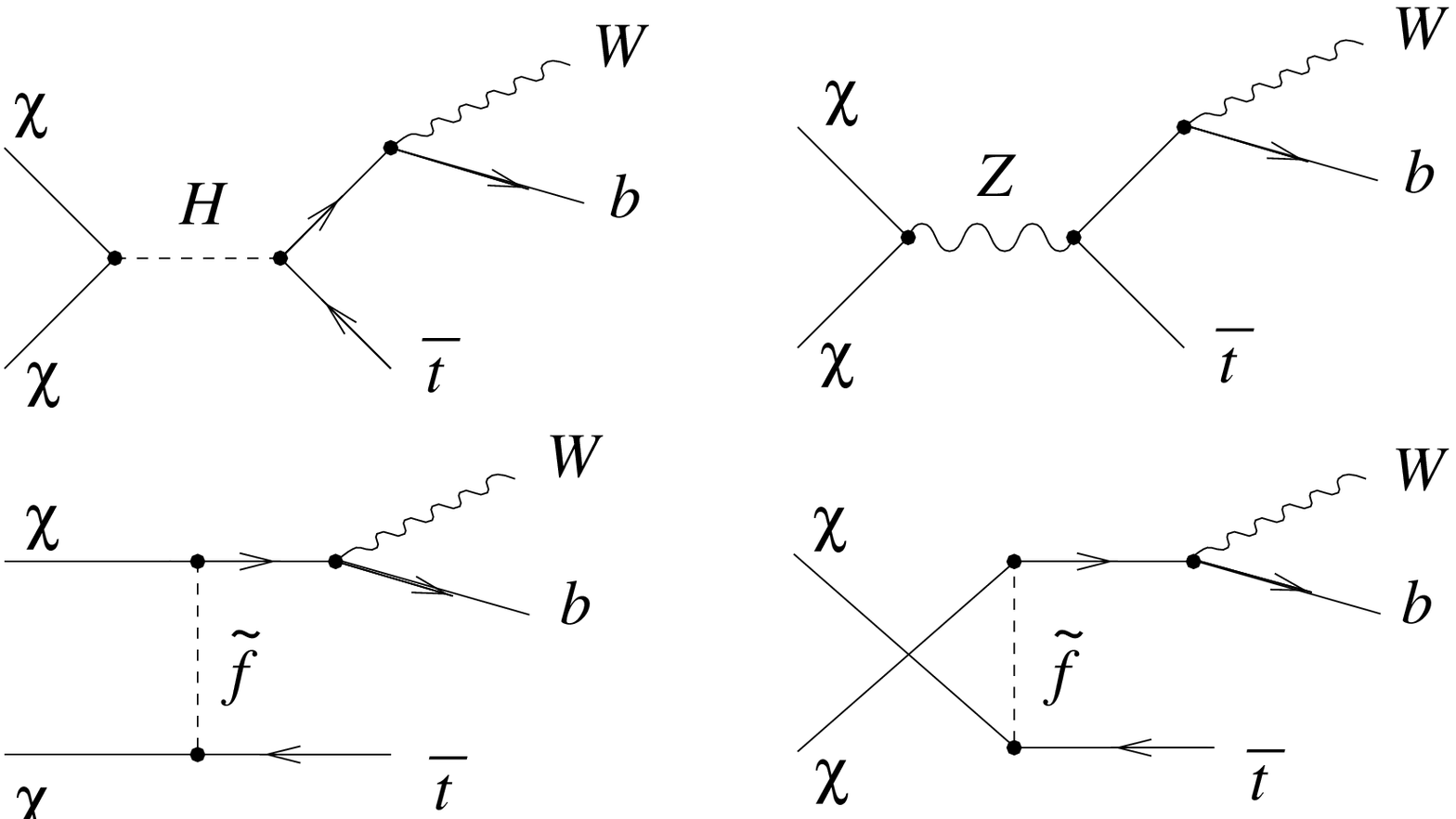,width=\linewidth}
\caption{Feynman diagrams for $\chi \chi \rightarrow t t^*$.}
\label{Feynman_twb}
\end{minipage}
\begin{minipage}[b]{0.45\linewidth}
\centering\epsfig{file=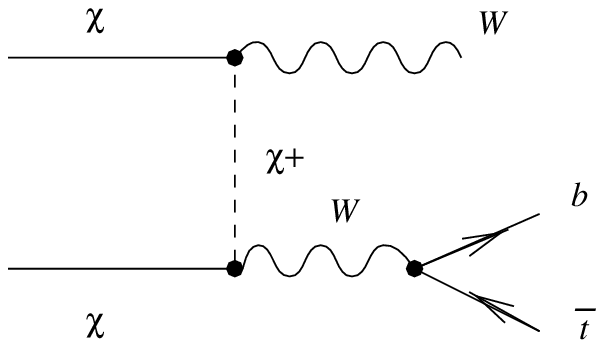,width=\linewidth}
\caption{neglected diagrams}
\label{Feynman_twb_omit}
\end{minipage}
\end{figure}

Below the top pair threshold, the neutralino may annihilate via a
virtual top quark: $\chi\chi \rightarrow t t^{*} \rightarrow t W b$.  
The Feynman diagrams for
this process are shown in Fig.~\ref{Feynman_twb}.  Like annihilation
to $t\bar t$ pairs, annihilation to this three-body final state
takes place via $s$-channel exchange of $Z^0$ and $A^0$
(pseudo-scalar Higgs bosons) and $t$- and $u$-channel exchange
of squarks.  Although there are additional diagrams for this
process, such as those shown in Fig.~\ref{Feynman_twb_omit}, these are
negligible for the gaugino because of the small coupling. On the other
hand, for higgsinos the gauge boson channel dominates and the $tt^{*}$
would be unimportant anyway. 
In the $v_{\rm rel} \to 0$ limit
\begin{equation}
\label{phase space}
     \sigma v_{\rm rel} = \frac{N_{c}}{128\pi^{3}}
     \int_{x_{6min}}^{x_{6max}}dx_{6}
     \int_{x_{4min}}^{x_{4max}}dx_{4}\frac{1}{4}|{\cal M}|^{2}\,,
\end{equation}
the amplitude is given by $M=M_{t} -M_{u}+M_{s}$.

\begin{figure*}[b]
\begin{minipage}[b]{0.4\linewidth}
\centering\epsfig{file=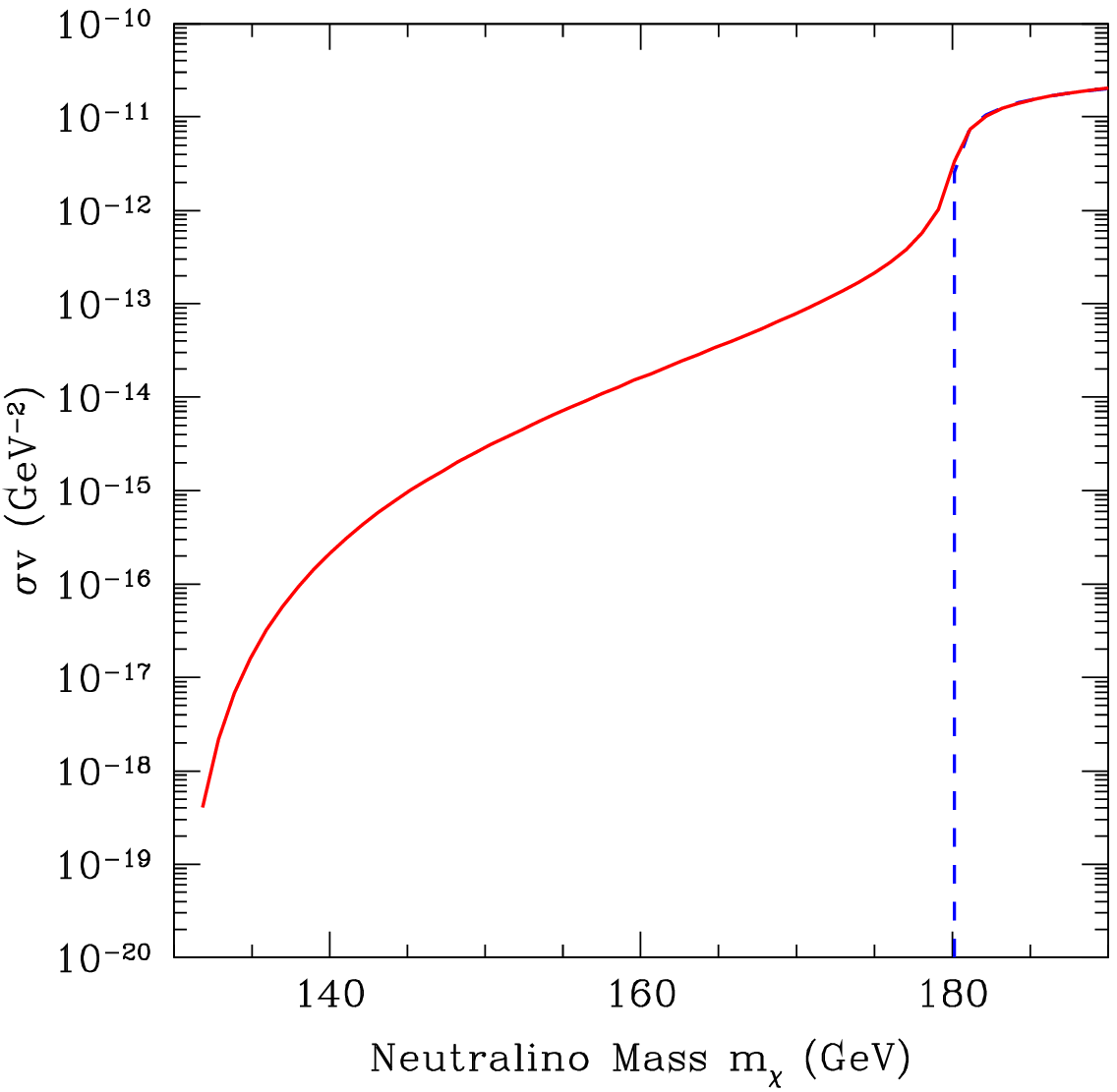,width=\linewidth}
\caption{total annihilation cross section.}
\label{twbcross}
\end{minipage}
\begin{minipage}[b]{0.4\linewidth}
\centering\epsfig{file=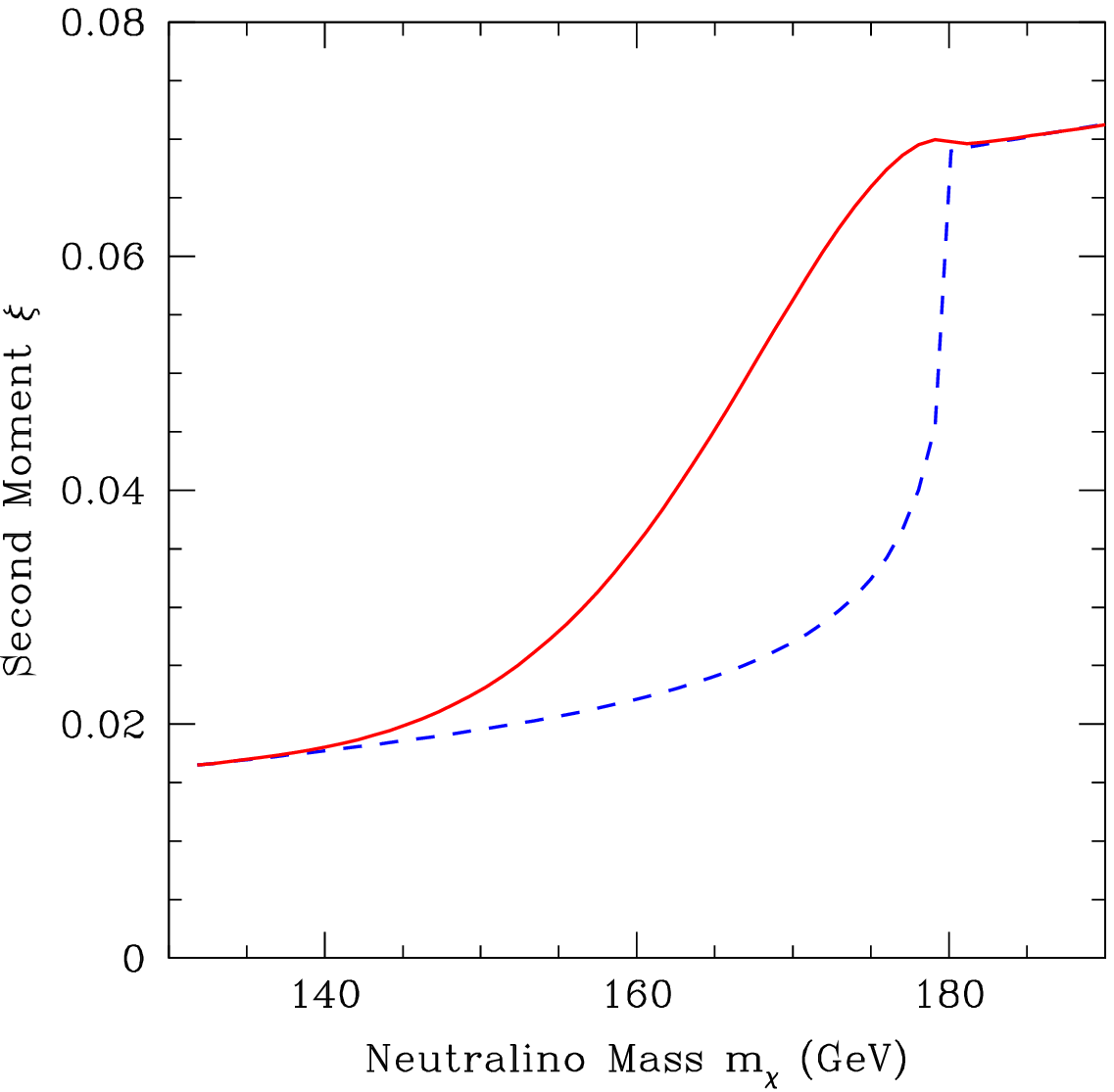,width=\linewidth}
\caption{Detection Rate for neutrinos from the core of Earth.}
\label{Nz2figureEarth}
\end{minipage}

{\it Dashed curves show two-body only result, solid curves
includes three-body contribution.}
\end{figure*}

The three-body cross section for a series of typical models are shown in 
Fig.~\ref{twbcross}. In these models the neutralino is primarily gaugino.
As expected, the three-body cross section 
approaches the two-body value above the top threshold 
(we take $m_{t}=180$ GeV). 
Below the top mass it is non-zero but drops quickly.
The flux of upward muons is proportional to the second moment of the 
neutrino energy spectrum weighted by branching ratios, which is given by
\begin{equation}
B_F \VEV{Nz^2}=\frac{3}{128\pi^{3}} \int dx_{4} dx_{6}|{\cal M}|^{2} 
\left(\VEV{Nz^2}_{t}x_{4}^{2} 
+\VEV{Nz^2}_{W} x_{5}^{2}
+\VEV{Nz^2}_{b} x_{6}^{2}\right),
\label{eqnxi}
\end{equation}
in the three-body case. Although the three-body cross section is small except
just below the $t\bar t$ threshold, its contribution to the the 
second moment, $B_F \VEV{Nz^2}$, may be important, as illustrated in 
Fig. ~\ref{Nz2figureEarth}. This is because the $\VEV{Nz^2}$ for
top quarks and $W$ bosons is significantly larger than that for the
light fermions \cite{neutrino}.

Below the $W^{+} W^{-}$ threshold, the neutralino can annihilate to a real
$W$ and a virtual $W$. The $W$-bosons then decay independently  
into a fermion pair $f\bar{f'}$, which can be $\tau\nu$,
$\mu\nu$, $e\nu$, $c s$, or $u d$.
About 10\% of these decay into a
muon (or anti-muon) and a muon anti-neutrino (or neutrino).

\begin{figure}[t]
\centering\epsfig{file=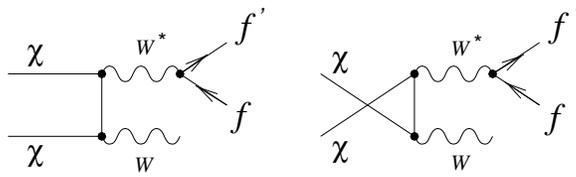,width=3in}
\caption{Feynman Diagrams for $\chi\chi \rightarrow W W^{*}$}\label{Feynman_ww}
\end{figure}

\begin{figure}[b]
\begin{minipage}[b]{0.4\linewidth}
\centering\epsfig{file=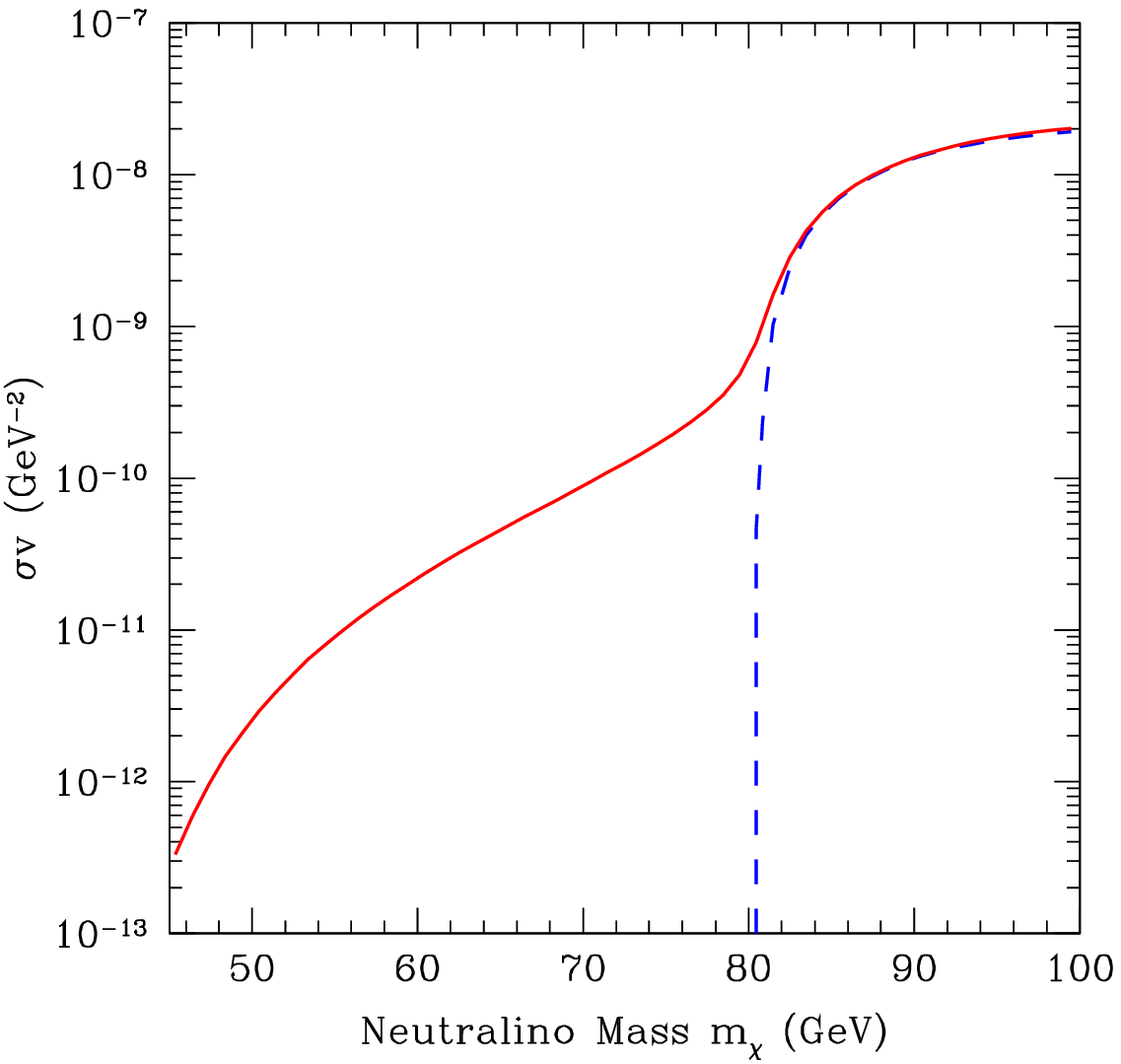,width=\linewidth}
\caption{total annihilation cross section.}
\label{W3cross}
\end{minipage}
\begin{minipage}[b]{0.4\linewidth}
\centering\epsfig{file=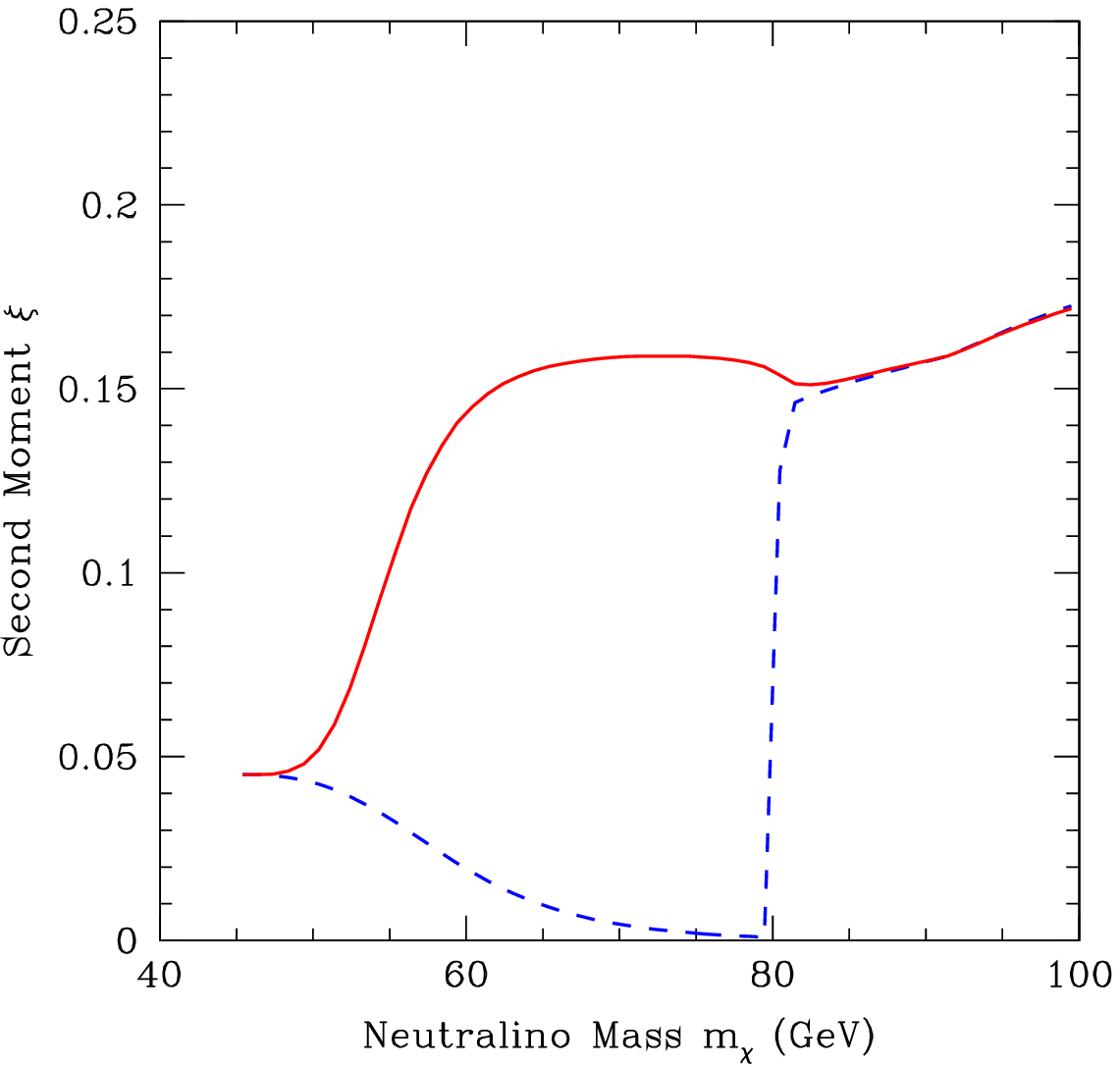,width=\linewidth}
\caption{Detection Rate for neutrinos from the core of Earth.
\label{W3Earth}}
\end{minipage}
\end{figure}

The $W W^*$ calculation is similar to the $tWb$ calculation. 
In the $v \to 0$ limit, only chargino exchange in the $t$ and $u$
channels shown in Fig.~\ref{Feynman_ww} are important.
Neutrinos can be produced either by the
virtual $W^* $ or by decay of the real $W$.
The neutrinos produced by decay of the muon or other fermions can be
neglected, and 
the real $W$ boson has a
probability $\Gamma_{W \to \mu\nu}$ to decay to a muon neutrino.
On the other hand this $W$ boson is produced in all $\chi\chi \to W
f\bar{f'}$ channels, and each of these channels has approximately
the same cross section, so the contribution has a weight of 
$n_{chan}\Gamma_{W \to \mu\nu} \approx 1$. The cross section and 
second moment for annihilation
in the Earth are shown in Fig.~\ref{W3cross} and Fig.~\ref{W3Earth}. 
Below $m_{W}/2$, the four-body channel might 
become significant, and may smooth the jump in much the same fashion
as the three-body channel does near the two-body channel threshold, but
we will not consider it here.

\section{Conclusions}

The contributions of these three-body channels are 
important only in a limited region of parameter space. However, they
may produce a
large effect. In fact, our calculation shows that although the cross
sections of these annihilations are significant only just below the 
two-body channel threshold, due to the high energy of the neutrinos
they produce, they can enhance the neutrino signal by many times and
actually dominate the neutrino signal far below the two-body threshold.  
Furthermore, the regions in question may be of particular interest. For
example, motivated by collider data, Kane and Wells proposed a light
higgsino \cite{light higgsino}, and recent DAMA results \cite{DAMA}
suggest a WIMP candidate with $m_{x} \simeq 60 $ GeV
 (but see \cite{EllisCosmo98}). 
There are also arguments that the
neutralino should be primarily gaugino with a mass somewhere
below but near the top-quark mass \cite{leszek}. 

There are many parameters in the minimal supersymmetric model. The results
shown in Figs.~\ref{Nz2figureEarth} and \ref{W3Earth}
 are of course model dependent, and these effects 
might be more or less important in models with different parameters.

\acknowledgments

We thank G. Jungman for help with the {\tt neutdriver} code.
This work was supported by D.O.E. Contract No. DEFG02-92-ER 40699, 
NASA NAG5-3091, and the Alfred P. Sloan Foundation.

\end{document}